\documentclass{aastex631}
\usepackage{bm}
\usepackage{subcaption}
\usepackage[T1]{fontenc}    
\usepackage{booktabs}       
\usepackage{threeparttable}
\usepackage{multirow}
\usepackage{amsmath}

\begin{document}

 \title{Simulation method of urban evacuation based on mesoscopic cellular automata\footnote{Received 4 January 2021; revised manuscript received 19 February 2021.  \\Acta Physica Sinica [In Chinese]}}

\author{Wei Lv}
\affiliation{China Research Center for Emergency Management\\
Wuhan University of Technology\\
Wuhan, China}
\affiliation{School of Safety Science and Emergency Management\\
Wuhan University of Technology\\
Wuhan, China}

\author{Jinghui Wang}
\affiliation{China Research Center for Emergency Management\\
Wuhan University of Technology\\
Wuhan, China}
\affiliation{School of Safety Science and Emergency Management\\
Wuhan University of Technology\\
Wuhan, China}

\author{Zhiming Fang}
\altaffiliation{{\url{zhmfang2015@163.com}}}
\affiliation{Business School\\
University of Shanghai for Science and Technology\\
Shanghai, China}

\author{Dun Mao}
\affiliation{China Research Center for Emergency Management\\
Wuhan University of Technology\\
Wuhan, China}

\affiliation{School of Safety Science and Emergency Management\\
Wuhan University of Technology\\
Wuhan, China}



\begin{abstract}

This study integrates pedestrian flow characteristics to formulate a mesoscopic cellular automata model tailored for simulating evacuations in large-scale scenarios. Departing from the conventional planar grid cell division, the model employs road cell segmentation, thereby physically enlarging the dimensions of individual cells. This augmentation accommodates an increased occupancy of individuals per cell, representing pedestrian flow parameters within each cell through state variables. The source loading cell facilitates the simulation of pedestrian behavior transitioning from buildings to roads during an actual evacuation event, while the unloading cell situated at the exit removes evacuees from the system. The continuity equation for state transitions comprehensively encapsulates the dynamics of pedestrians throughout the evacuation process. Potential challenges in actual evacuation processes are identified through the simulation, offering valuable insights for improvement. This research aims to contribute to a more effective and informed approach to evacuation planning and management.

\end{abstract}



\section{Introduction} 
\label{section1}

In recent years, with the expansion of urban and the rapid increase in population in our country, coupled with the ongoing deterioration of the global climate environment, frequent natural disasters such as earthquakes, tsunamis, floods, as well as accidents and disasters like fires, explosions, and hazardous material leaks, have caused significant loss of life and economic damage to many cities. Urban public safety is facing tremendous pressure and challenges. When cities experience major disaster incidents, the rapid evacuation of the population within the affected area to safe zones is a crucial measure to ensure the safety of people's lives. The issue of regional evacuation in urban areas is gradually becoming a new research focus in the field of evacuation.

Unlike traditional scenarios of building fire evacuation, regional evacuation scenarios involve larger spatial extents, longer evacuation distances, and evacuation facilities that depend more on the region's road infrastructure. Therefore, conducting large-scale evacuation drills to analyze the feasibility and effectiveness of regional emergency evacuation plans is evidently impractical. It is necessary to employ simulation techniques based on evacuation models for research. By simulating the movement of individuals through computer modeling, results such as evacuation time, efficiency, personnel density, and flow rate can be calculated. These data are essential for assessing evacuation risks, optimizing evacuation plans, and addressing practical concerns.

In the field of personnel evacuation modeling, cellular automata model, due to its efficiency and ease of implementation, has been widely utilized in simulating the evacuation of personnel in architectural spaces. Similarly, it possesses the capability to simulate emergency evacuations in urban areas.

In the realm of cellular automata evacuation models, many scholars have initiated research early on and conducted extensive expansion model studies in various evacuation scenarios. \citet{miyagawa2020cellular} established a multi-grid cellular automata model for crowd evacuation, investigating the impact of lateral movement and turning behavior on the evacuation process. \citet{ji2018cellular} changing the cellular shape from rectangular to triangular, increased the number of movement directions of pedestrians in the model from 8 to 14. \citet{maniccam2003traffic} transformed the cellular shape from rectangular to hexagonal and studied the variation in critical density within the cells. \citet{leng2014extended} developed a field cellular model based on regular hexagons, studying dynamic characteristics of pedestrians in corridors. \citet{kim2018modeling} expanding the cellular automata model, simulated the influence of differently proportioned disabled populations on channel evacuation. \citet{hanisch2003online} employed cellular transport models from traffic research to study pedestrian flow in public spaces, and \citet{bandini2017collision} proposed a multi-scale modeling approach based on cellular automata for simulating ferry transportation scenes in Staten Island, New York. \citet{lammel2015model} established an event-driven cellular automata model, simulating single and bidirectional pedestrian flow evacuation scenarios. \citet{kaji2017cellular} combining multi-grid cellular automata models and static field models, studied the flow characteristics of fluids in laminar flow. \citet{shi2018novel} introduced a mesoscopic evacuation model and developed a new static field algorithm for modeling personnel evacuation at the mesoscopic scale.

Some scholars have primarily focused on micro-scale cellular automata evacuation models within building interiors. \citet{hu2014cellular} developed a three-dimensional cellular automata model to study pedestrian flow characteristics on stairs. \citet{lei2015simulation} used cellular automata models to study the mechanisms of congestion-induced injuries in crowded pedestrian areas. \citet{Ren} established a cellular automata model considering pedestrians' walking tendency characteristics and analyzed pedestrian flow features using complex network theory. \citet{Xu2018} introduced a cellular automata model considering detour behavior by introducing a pedestrian perception range parameter. \citet{Jing2018} considering pedestrians' scenario-based movement direction and expected velocity rules, expanded the field strength model. \citet{cao2015multi} established a multi-grid personnel evacuation model in low visibility environments, studying the impact of exit width, exit quantity, and initial density on evacuation time.

Summarizing the current research status both domestically and internationally reveals that existing cellular automata evacuation models are primarily sophisticated grid models at the microscopic level. They are mainly applied to two-dimensional plane scenarios, with evacuation spaces predominantly confined to small-scale architectural or local areas. However, cellular automata models suitable for simulating large-scale urban area evacuations are still immature. This is mainly due to the high-precision characteristics of microscopic cellular automata models, which impose significant constraints on computational efficiency when expanding to large-scale spaces.
Therefore, to explore the applicability of cellular automata evacuation models in large-scale spaces, this paper proposes a method for urban area emergency evacuation simulation based on mesoscopic cellular automata. This involves partitioning the road network of the evacuation area into microscopic cells, loading and unloading personnel flow on specific cells to facilitate evacuation generation and completion, and transferring macroscopic state variables on road cells to realize the evacuation process. The aim is to provide scientific support for the demonstration of large-scale urban emergency evacuation plans.

\section{Model} \label{section2}
\subsection{Microscopic cellular automata evacuation model} \label{subsection2.1}

\begin{figure}[ht!]
\centering
\includegraphics[scale=0.5]{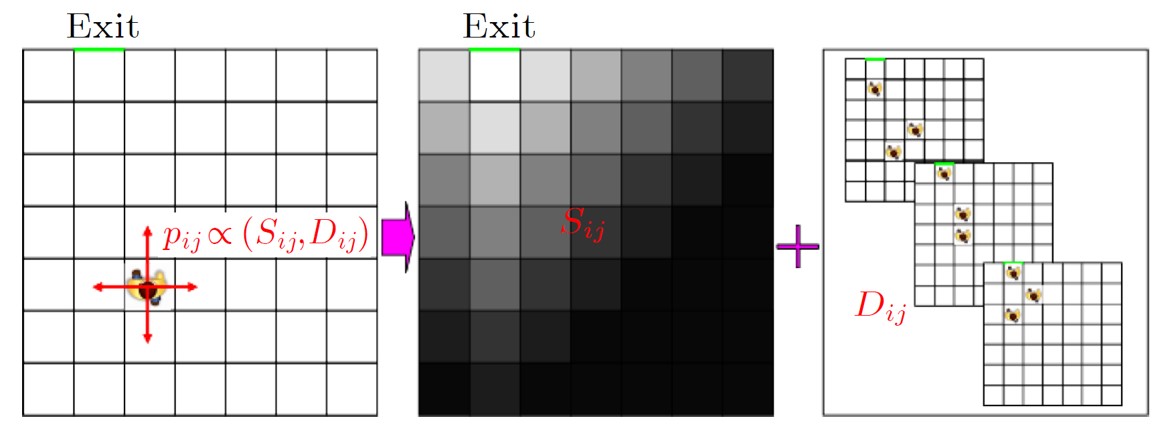}
\caption{Sketch map of the movement process in FF model. 
\label{fig1}}
\end{figure}

Presently, conventional cellular automata evacuation models predominantly constitute microscopic grid-based evacuation models. This category encompasses spatiotemporal discrete models that alter the system's state according to predefined rules, including biased random walk models with no backtracking \citep{muramatsu1999jamming}, field models \citep{burstedde2001simulation}, multi-grid models \citep{xu2008discretization} and k-nearest neighbor models\citep{ma2010k}, etc. The core concept of these models revolves around effecting changes in the spatial state of evacuated individuals through their migration on cellular automata. Consequently, the crux of these models lies in the computation of migration probabilities between cells.
Taking the classical field floor model (FF) as an illustrative example, it incorporates the consideration of the impact of the system's structure or architectural spatial configuration on evacuees through static field. Additionally, it considers the guiding effects of crowd movement in the system through dynamic field. Subsequently, based on Formula 1, it calculates the transfer probabilities of evacuees on each cell under the combined influence of the static and dynamic field. These probabilities are then utilized to update the spatial positions of evacuees within the system, facilitating the simulation of dynamic evacuation processes, as depicted in Fig.\ref{fig1}.

\begin{equation}\label{1}
{{p}_{ij}}={{(\sum{{{p}_{ij}}})}^{-1}}\exp ({{k}_{D}}{{D}_{ij}})\exp ({{k}_{S}}{{S}_{ij}})(1-{{n}_{ij}}){{\xi }_{ij}}
\end{equation}

It is evident that cellular automata evacuation models, represented by the field model, are typical microscopic-scale models. As the cellular scale increases, the efficiency of simulation computations will significantly decrease. Consequently, these models face challenges in their applicability to large-scale spatial urban regional emergency evacuation simulation studies.

\subsection{Mesoscopic cellular automata evacuation model} \label{subsection2.2}
\begin{figure}[ht!]
\centering
\includegraphics[scale=0.25]{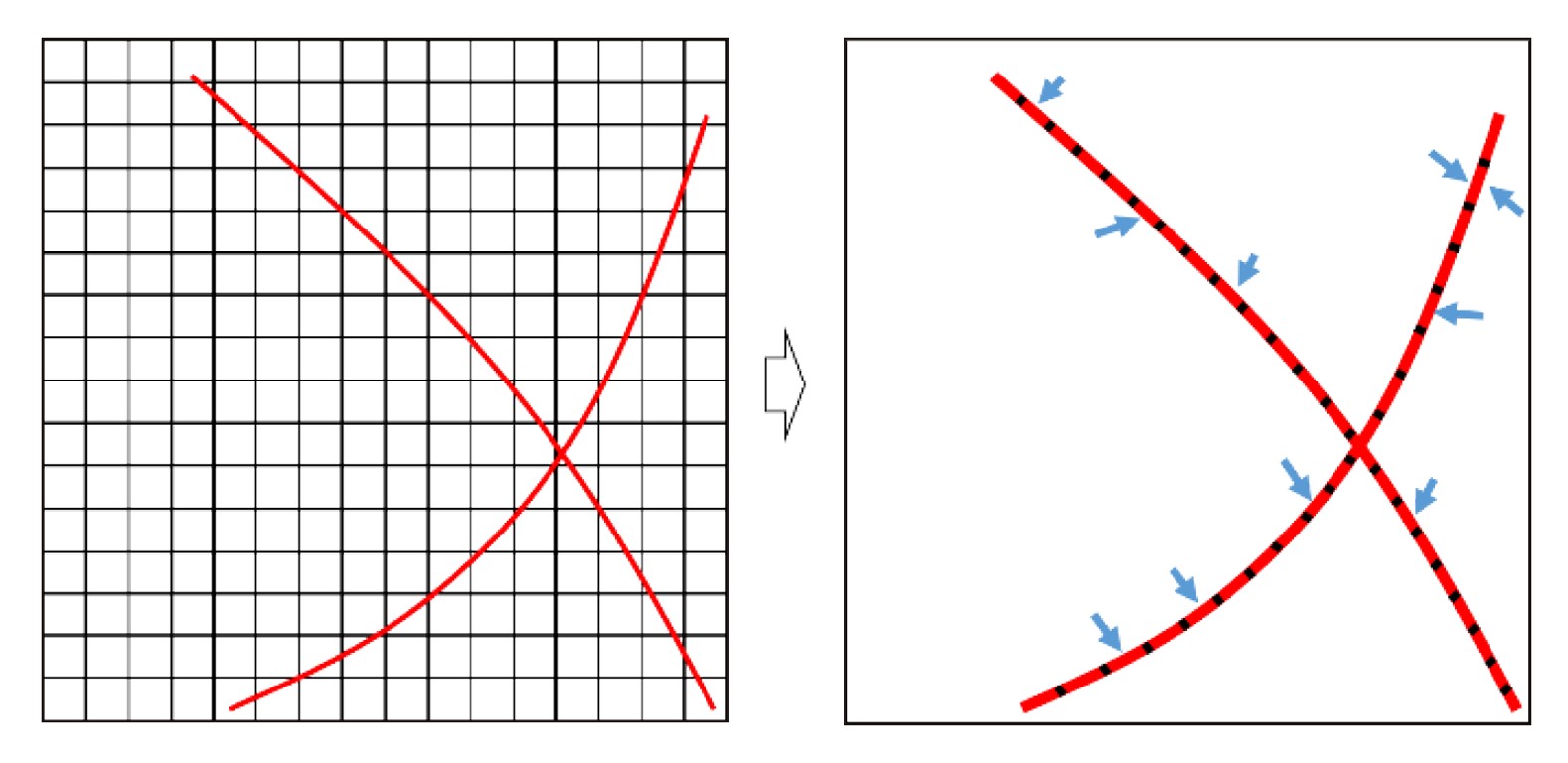}
\caption{Change from plane mesh to road cell segmentation. 
\label{fig2}}
\end{figure}

\begin{figure}[ht!]
\centering
\includegraphics[scale=0.25]{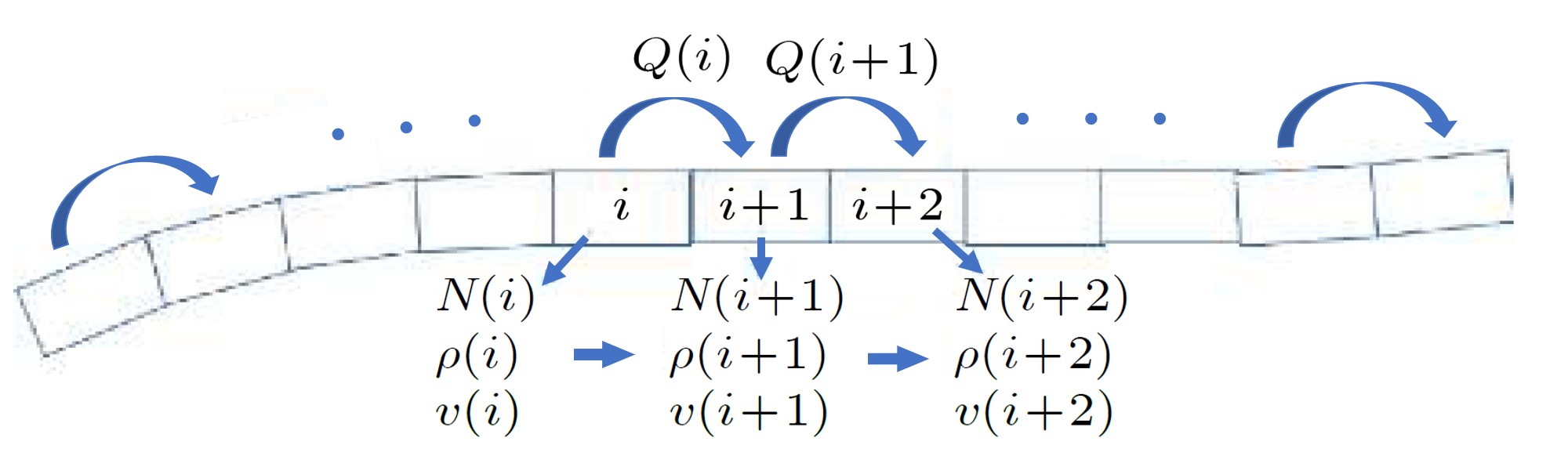}
\caption{The update of macro-state variable between road cells. 
\label{fig3}}
\end{figure}

To facilitate the application of cellular automata model in urban regional emergency evacuation simulations, the fundamental concept of the classical cellular automata evacuation model, which involves updating the system state through the migration of evacuees between cells, must be maintained. However, at the model scale, a transition from the microscopic to mesoscopic level is imperative.
The term "mesoscopic" encompasses two aspects: firstly, in terms of spatial cell division, the conventional planar grid division method is abandoned in favor of a road cell segmentation approach. Spatial discretization is limited to the evacuation road network, and the transfer of individuals from other areas to the roads is manifested through the source loading of specific cells in the evacuation road network, as illustrated in Fig.\ref{fig2}. Secondly, in terms of evacuee cell migration, the conventional approach of individuals moving based on transfer probabilities is abandoned. Instead, the evacuation process is directly realized through the update of macroscopic state variables corresponding to cells, including evacuees count, population density, evacuation speed, etc., as shown in Fig.\ref{fig3}.

Following the outlined principles, the fundamental procedure for constructing and simulating a mesoscopic cellular automata model for urban regional emergency evacuation is as follows:

\centerline{}
1. Define the Boundary of the Evacuation Region:

• Determine the scope of the urban regional area under investigation.

• Extract the evacuation road network within this defined boundary.

\centerline{}
2. Cell Segmentation of Evacuation Roads:

• Divide the evacuation roads into cells.

• Identify cells designated for source loading and cells corresponding to evacuation exits.

\centerline{}
3. Distance Calculation and Exit Assignment:

• Compute the distances between each road cell and each exit cell.

• Utilize a proximity-based exit selection principle to assign each road cell to the exit cell it points towards.

\centerline{}
4. Establish Macroscopic State Update Rules:

• Develop rules for updating macroscopic state variables of road cells based on principles of traffic flow conservation, capacity constraints, and evacuation traffic flow patterns.

\centerline{}
5. Simulation Conditions Setup:

• Define simulation conditions, including initial parameters and constraints.

• Initiate calculations and updates for the state values of road network cells.

• Iterate through calculations and updates until the population within the evacuation road network decreases to a critical threshold, signifying the completion of the evacuation process.

\begin{figure}[ht!]
\centering
\includegraphics[scale=0.25]{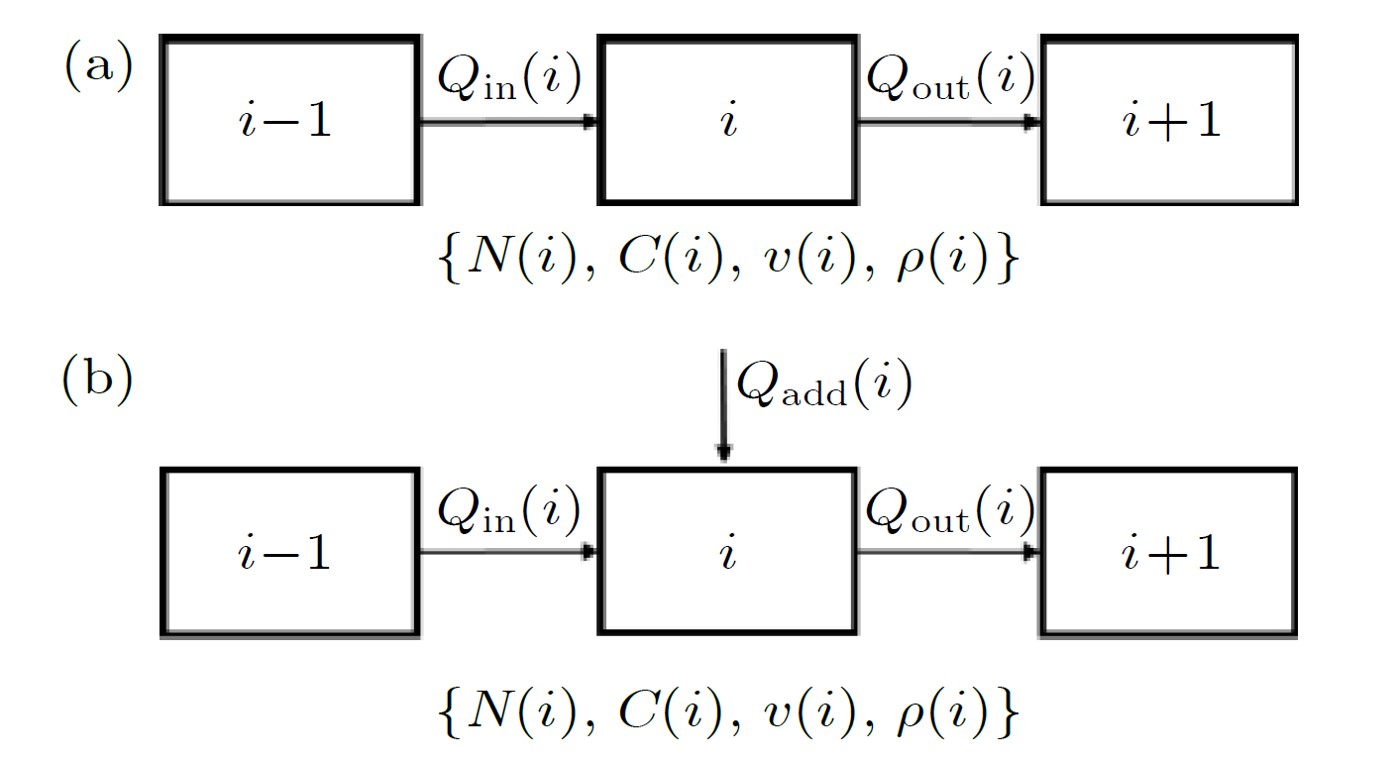}
\captionsetup{singlelinecheck=false, justification=centering}
\caption{The update rule of macro-state variable between the evacuation road cells:\\(a) Without load source; (b) with load source.
\label{fig4}}
\end{figure}

\centerline{}
In conjunction with Fig.\ref{fig4}, within the mesoscopic cellular automata evacuation model, the fundamental morphological forms of road cells essentially comprise two types. One type is the ordinary cell without external personnel loading, as illustrated in Fig.\ref{fig4}(a). The other type is the specific cell with external source loading, as depicted in \ref{fig4}(b). Regardless of the cell type, the rules followed are similar, and both must adhere to the principles of traffic flow conservation, capacity constraints, and evacuation traffic flow patterns.

In terms of traffic flow conservation, the update of the personnel quantity \(N(i)\) within cell \(i\) must adhere to the following equation:

\begin{equation}\label{2}
{{N}_{t+\Delta t}}(i)={{N}_{t}}(i)+{{Q}_{in,\Delta t}}(i)-{{Q}_{out,\Delta t}}(i)
\end{equation}

Regarding capacity constraints, the personnel quantity \(N_i(t)\) within cell \(i\) must not exceed the maximum capacity \(C_i\) of cell \(i\):

\begin{equation}\label{3}
{{N}_{t+\Delta t}}(i)={{N}_{t}}(i)+{{Q}_{in,\Delta t}}(i)-{{Q}_{out,\Delta t}}(i)
\end{equation}

Concerning the dynamics of pedestrian flow, the flow between cell \(i\) and its adjacent cells must adhere to the flow-density relationship:

\begin{equation}\label{4}
{{Q}_{in,\Delta t}}(i)={{\rho }_{t}}(i-1)\cdot {{v}_{t}}(i-1)\cdot w\cdot \Delta t
\end{equation}

\begin{equation}\label{5}
{{Q}_{out,\Delta t}}(i)={{\rho }_{t}}(i)\cdot {{v}_{t}}(i)\cdot w\cdot \Delta t
\end{equation}

In the aforementioned formula, \(N_i(t)\) represents the quantity of individuals within cell \(i\) at time \(t\), measured in persons; \(Q_i(t)\) signifies the quantity of individuals entering and exiting cell \(i\) during the time interval \(\Delta t\), measured in persons; \(C_i\) denotes the maximum capacity of individuals that cell \(i\) can accommodate, measured in persons; \(\rho_i(t)\) and \(V_i(t)\) respectively represent the density (measured in persons per square meter) and flow velocity (measured in meters per second) of individuals within cell \(i\) at time \(t\); \(w\) signifies the width of the cell, measured in meters.

Based on the three fundamental rules outlined above, the macroscopic state variable update equation for unloaded source cells is established as follows:

\begin{equation}\label{6}
\left\{ \begin{aligned}
  & {{Q}_{in,\Delta t}}(i)=\min \{{{\rho }_{t}}(i-1)\cdot {{v}_{t}}(i-1)\cdot w\cdot \Delta t,C(i)-{{N}_{t}}(i)\} \\ 
 & {{Q}_{out,\Delta t}}(i)=\min \{{{\rho }_{t}}(i)\cdot {{v}_{t}}(i)\cdot w\cdot \Delta t,C(i+1)-{{N}_{t}}(i+1)\} \\ 
 & {{N}_{t+\Delta t}}(i)=\min \{{{N}_{t}}(i)+{{Q}_{in,\Delta t}}(i)-{{Q}_{out,\Delta t}}(i),C(i)\}
\end{aligned} \right.
\end{equation}

The macroscopic state variable equation for loaded source cells is presented as follows:

\begin{equation}\label{7}
\left\{ \begin{aligned}
& if\thinspace\thinspace\begin{matrix}
   {}  \\
\end{matrix}{{Q}_{in,\Delta t}}(i)+{{Q}_{add,\Delta t}}(i)\le C(i)-{{N}_{t}}(i) \\ 
 & \begin{matrix}
   {}  \\
\end{matrix}{{Q}_{in,\Delta t}}(i)={{\rho }_{t}}(i-1)\cdot {{v}_{t}}(i-1)\cdot w\cdot \Delta t \\ 
 & \begin{matrix}
   {}  \\
\end{matrix}{{Q}_{add,\Delta t}}(i)={{Q}_{add,\Delta t}}(i) \\ 
 & if\thinspace\thinspace\begin{matrix}
   {}  \\
\end{matrix}{{Q}_{in,\Delta t}}(i)+{{Q}_{add,\Delta t}}(i)>C(i)-{{N}_{t}}(i) \\ 
 & \begin{matrix}
   {}  \\
\end{matrix}{{Q}_{in,\Delta t}}(i)=\alpha \cdot [C(i)-{{N}_{t}}(i)] \\ 
 & \begin{matrix}
   {}  \\
\end{matrix}{{Q}_{add,\Delta t}}(i)=(1-\alpha )\cdot [C(i)-{{N}_{t}}(i)] \\ 
 & \begin{matrix}
   {}  \\
\end{matrix}\alpha ={{Q}_{in,\Delta t}}(i)/[{{Q}_{in,\Delta t}}(i)+{{Q}_{add,\Delta t}}(i)] \\ 
 & {{Q}_{out,\Delta t}}(i)=\min \{{{\rho }_{t}}(i)\cdot {{v}_{t}}(i)\cdot w\cdot \Delta t,C(i+1)-{{N}_{t}}(i+1)\} \\ 
 & {{N}_{t+\Delta t}}(i)=\min \{{{N}_{t}}(i)+{{Q}_{in,\Delta t}}(i)+{{Q}_{add,\Delta t}}(i)-{{Q}_{out,\Delta t}}(i),C(i)\}
\end{aligned} \right.
\end{equation}

Within the state update equation, the flow velocity, capacity, and density within cell \(i\) can be computed using the following formulas:

\begin{equation}\label{8}
{{v}_{t}}(i)={{v}_{f}}\cdot \exp [-{{\rho }_{t}}(i)/{{\rho }_{m}}]
\end{equation}

\begin{equation}\label{9}
{{\rho }_{t}}(i)={{N}_{t}}(i)/({{l}_{i}}\cdot w)
\end{equation}

\begin{equation}\label{10}
C(i)={{l}_{i}}\cdot w\cdot {{\rho }_{m}}
\end{equation}

Where, \( \rho_c \) represents the critical density, denoting the threshold at which individuals can only proceed at an exceedingly minimal pace. Typically, the critical density is set at 5 persons per square meter (\(5 \, \text{ped/m}^2\)). \(L_i\) signifies the length of cell \(i\), measured in meters, and \(V_f\) denotes the free evacuation velocity, measured in meters per second.

\section{Modeling Urban Regional Evacuation Based on Mesoscopic Cellular Automata} \label{section3}

\subsection{Study area} \label{subsection3.1}

\begin{figure}[ht!]
\centering
\includegraphics[scale=0.3]{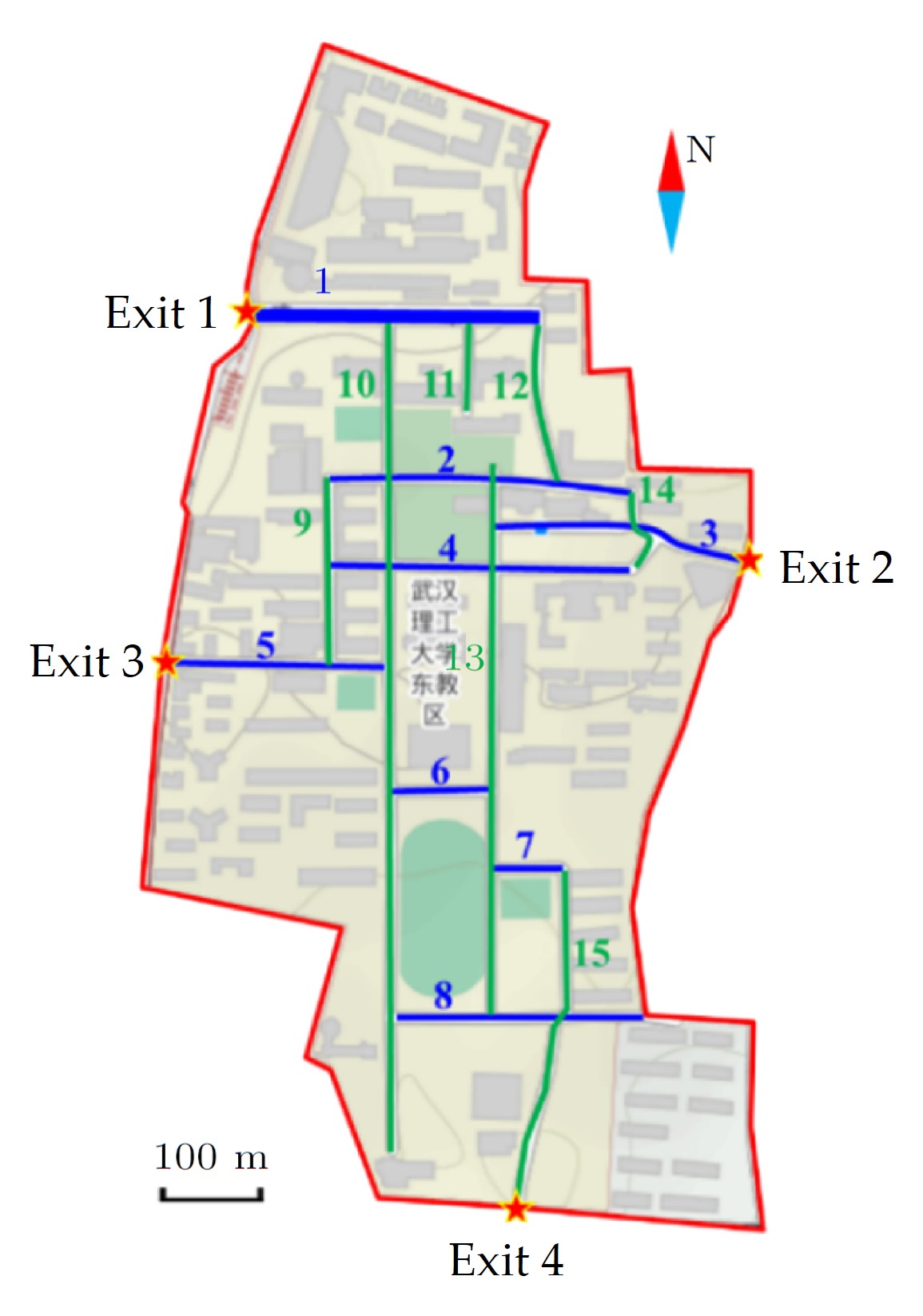}
\caption{The spatial scope and boundary of the study area. 
\label{fig5}}
\end{figure}

Given the expansive spatial scope and high population density characteristic of university campuses, which make them well-suited for conducting large-scale regional evacuation simulation studies, this study selects a university campus area near the Jianding Street in the Hongshan District of Wuhan City, as depicted in Fig.\ref{fig5}. The feasibility study for regional emergency evacuation simulation is conducted using the mesoscopic cellular automata evacuation model proposed earlier. The selected campus area spans approximately 675 acres (450,000 m²) and includes 2 academic buildings, 1 library, 10 student dormitories, and 39 faculty dormitories. According to surveys, the maximum capacity of all buildings is around 16,838 individuals, while the actual population on regular teaching days is approximately 8,200 individuals. Furthermore, the road network within the area consists of 8 east-west roads (numbered 1-8) and 7 north-south roads (numbered 9-15), with 4 exits (marked with asterisks) leading to external boundaries.

\subsection{Road network segmentation}\label{subsection3.2}

\begin{figure}[ht!]
\centering
\includegraphics[scale=0.25]{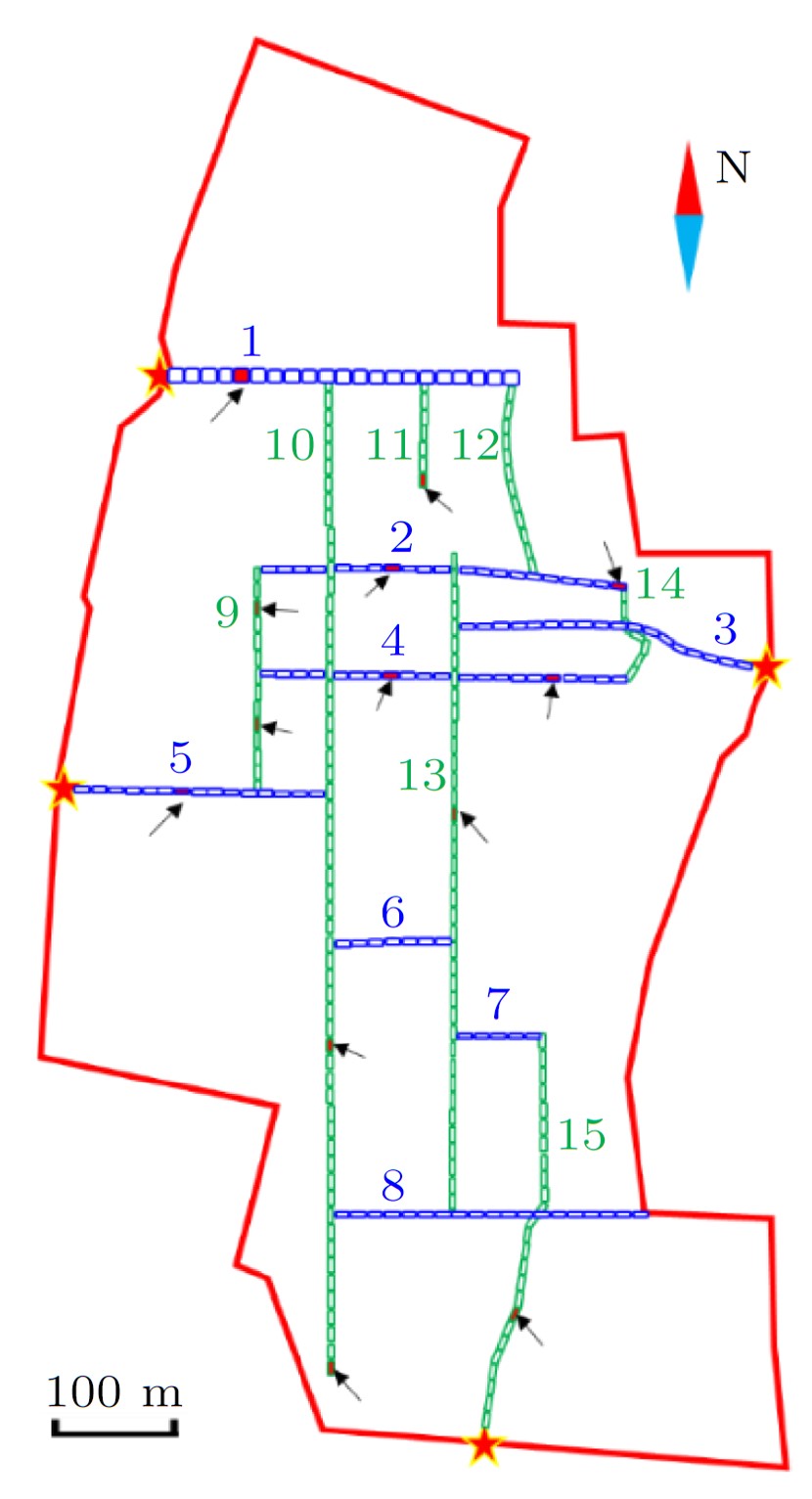}
\caption{The road cell segmentation of the road network. 
\label{fig6}}
\end{figure}

Along the directional alignment of the road network, with a fundamental cell size of 10 meters in length and 6 meters in width, the road network is partitioned into cellular units. Considering the locations of academic buildings, student dormitories, the library, and faculty-student residences, cells proximate to these densely populated areas are selected as evacuation source-loading cells (cells indicated by arrows), while boundary exit cells are chosen as evacuation exit cells (cells marked by asterisks), as illustrated in Fig.\ref{fig6}.

Furthermore, following the numbering of roads, cellular coding is sequentially applied to each road, starting from one end (west or north) and progressing towards the other end (east or south). The encoded cells are represented as \(Cell_{i,j}\), where \(i\) denotes the road number, and \(j\) represents the cell's position within the \(i\)-th road. The statistical information for the road cell encoding in the research area is presented in Tab.\ref{table1}.

\begin{table}[]
\caption{Statistical information of regional road cellular coding.}
\centering
\resizebox{.75\columnwidth}{!}{%
\centering
\begin{tabular}{lllllll}
\toprule
\textbf{Road ID} & \textbf{Length/m} & \textbf{Width/m} & \textbf{Number of Cells} & \textbf{Source Loading Cells} & \textbf{Exit Cells} \\ \midrule
1       & 280      & 12      & 28                & Cell$_{1,8}$              & Cell$_{1,1}$    \\
2       & 290      & 6       & 29                & Cell$_{2,9}$, Cell$_{2,29}$    & /          \\
3       & 230      & 6       & 23                & /                    & Cell$_{3,23}$   \\
4       & 290      & 6       & 29                & Cell$_{4,9}$, Cell$_{4,17}$    & /          \\
5       & 200      & 6       & 20                & Cell$_{5,9}$              & Cell$_{5,1}$    \\
6       & 90       & 6       & 9                 & /                    & /          \\
7       & 68       & 6       & 7                 & /                    & /          \\
8       & 250      & 6       & 25                & /                    & /          \\
9       & 180      & 6       & 18                & Cell$_{9,5}$, Cell$_{9,13}$    & /          \\
10      & 795      & 6       & 80                & Cell$_{10,59}$, Cell$_{10,80}$  & /          \\
11      & 90       & 6       & 9                 & Cell$_{11,9}$             & /          \\
12      & 148      & 6       & 15                & /                    & /          \\
13      & 530      & 6       & 53                & Cell$_{13,21}$            & /          \\
14      & 70       & 6       & 7                 & /                    & /          \\
15      & 326      & 6       & 33                & Cell$_{15,22}$            & Cell$_{15,33}$  \\ \bottomrule
\end{tabular}
}
\label{table1}
\end{table}

\subsection{Evacuation subnetworks division\label{subsection3.3}}

\begin{figure}[ht!]
\centering
\includegraphics[scale=0.6]{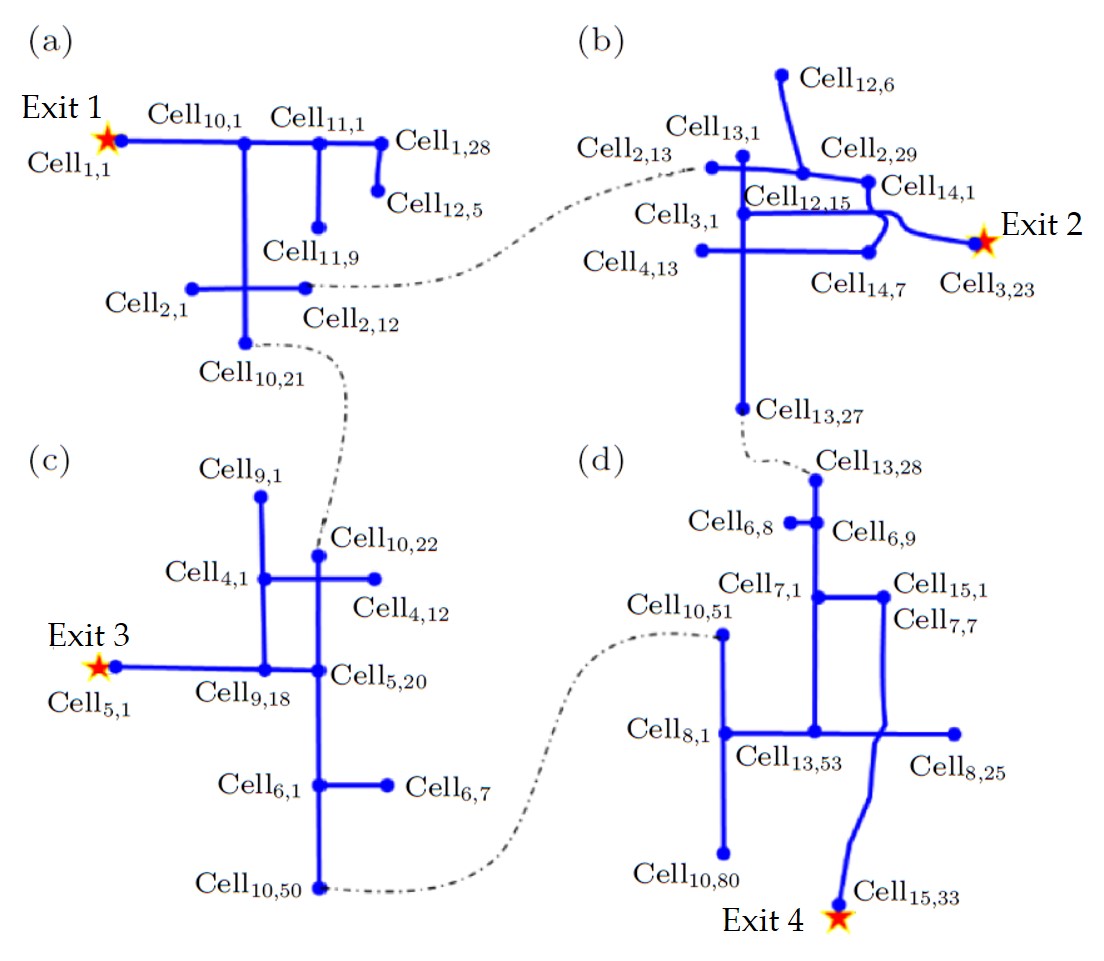}
\caption{Evacuation range of the roads covered by each exit. 
\label{fig7}}
\end{figure}

In this study, four exits are respectively designated as Exit 1, Exit 2, Exit 3, and Exit 4. By computing the distance between each road cell and the four designated exit cells and selecting the closest exit as the target exit for evacuation, the coverage range of road cells for the four exits is determined, as illustrated in Fig.\ref{fig7}.

\section{Simulation and results analysis\label{section4}}

\subsection{Scenario and parameter configuration\label{subsection4.1}}

The simulation scenario involves the rapid evacuation of the campus population to designated safe areas in the event of a significant disaster. Assuming a total evacuation requirement of 8200 individuals distributed across 13 source-loaded cells near instructional buildings, the library, and dormitory structures, all individuals promptly receive evacuation directives. Subsequently, they proceed to evacuate via walking from the source-loaded cells to the evacuation network. Participants enter the source-loaded cells on a first-come, first-served basis within capacity constraints, with no consideration given to the time it takes for individuals to traverse from their respective buildings to the source-loaded cells. The parameter values involved in the model are presented in Tab.\ref{table2}.

\begin{table}[!htb]\centering
\caption{Model parameters.}
\resizebox{.5\columnwidth}{!}{
\begin{tabular}{lll}
\hline
\textbf{Parameters} & \textbf{Meaning} & \textbf{Value} \\
\hline
\(l\) & Length of road cell & \(10 \, \text{m}\) \\
\(w\) & Width of road cell & \(6 \, \text{m}\) \\
\(v_f\) & Free evacuation speed & \(1.5 \, \text{m/s}\) \\
\(\Delta t\) & Timestep for calculation & \(1 \, \text{s}\) \\
\(\rho_m\) & Congestion critical density & \(5 \, \text{per/m}^2\) \\
\hline
\end{tabular}
}
\label{table2}
\end{table}

\begin{figure}[ht!]
\centering
\includegraphics[scale=0.25]{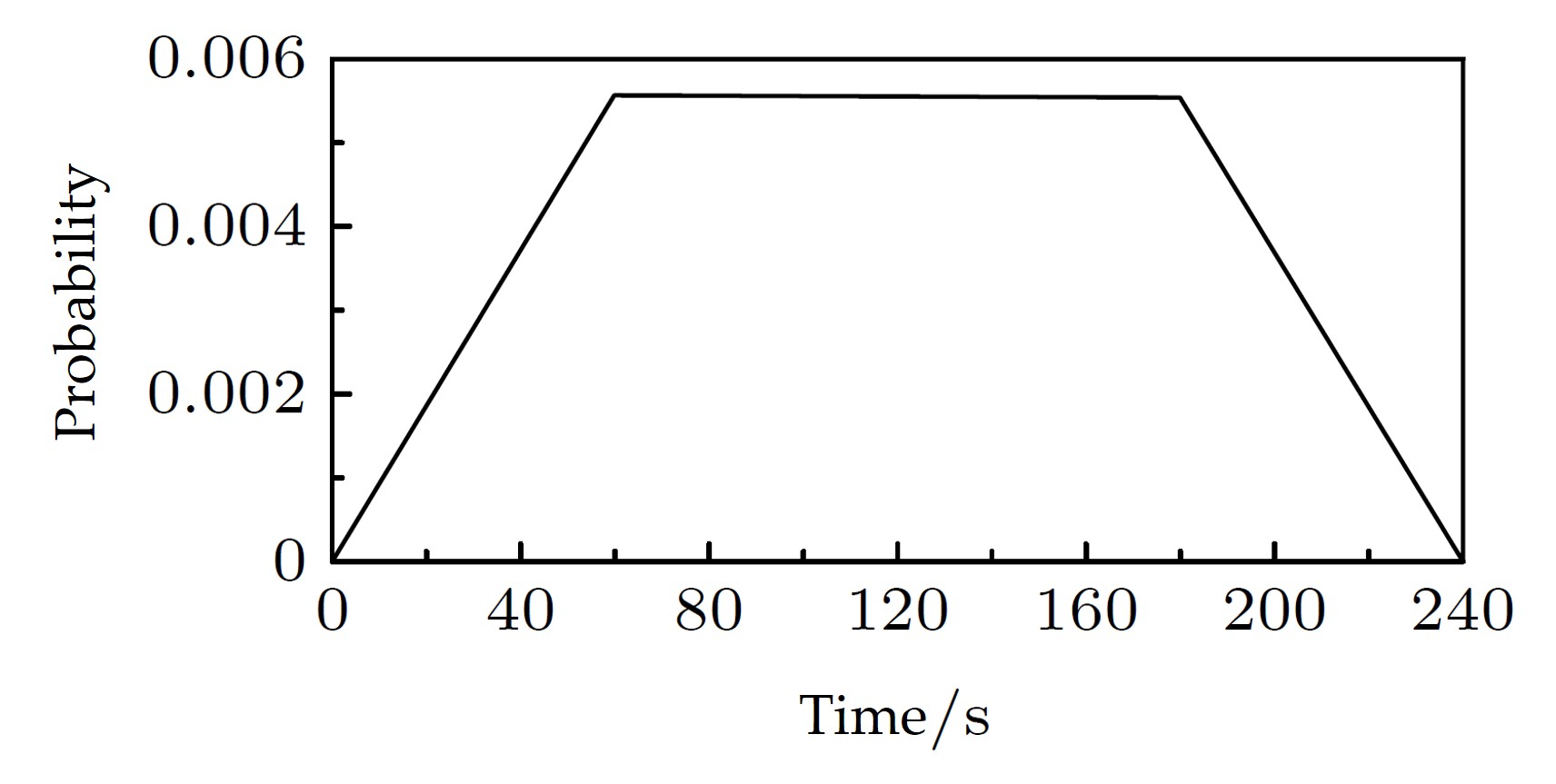}
\caption{Probability density distribution figure of  source loading cell. 
\label{fig8}}
\end{figure}

Based on empirical investigations, the loading capacity of each source-loaded cell was determined, yielding the actual loading capacities for each cell as presented in Tab.\ref{table3}. Furthermore, employing the evacuation subnetwork delineated in Fig.\ref{fig7}, the total evacuation counts for each exit during the evacuation process were derived and are detailed in Tab.\ref{table4}. Considering bottleneck effects during the evacuation from buildings, the loading probability for source-loaded cells was configured as a trapezoidal probability density distribution, as illustrated in Fig.\ref{fig8}. The loading quantity for each source-loaded cell in each timestep is the product of the loading probability for that timestep and the total loading capacity, respectively. The total duration for each source-loaded cell, from the initiation of loading to completion, spans 240 timesteps. Commencing at the onset of the simulation, personnel loading commences for each source-loaded cell and concludes at the 240th timestep. Subsequently, pedestrians who have been loaded move along the pathways to the exit cell, exiting the campus. The simulation is deemed concluded when the remaining personnel drop below 0.5 individuals, signifying the complete evacuation of all individuals from the campus.

\begin{table}[]
\caption{The number of pedestrians loaded by
each source loading cell.}
\centering
\begin{tabular}{ll}
\toprule
\textbf{Cell index} & \textbf{Total evacuees} \\ \midrule
\centering Cell$_{1,8}$ & 500 \\
\centering Cell$_{2,9}$ & 500 \\
\centering Cell$_{2,29}$ & 500 \\
\centering Cell$_{4,9}$ & 700 \\
\centering Cell$_{4,17}$ & 500 \\
\centering Cell$_{5,9}$ & 900 \\
\centering Cell$_{9,5}$ & 800 \\
\centering Cell$_{9,13}$ & 800 \\
\centering Cell$_{10,59}$ & 600 \\
\centering Cell$_{10,80}$ & 600 \\
\centering Cell$_{11,9}$ & 600 \\
\centering Cell$_{13,21}$ & 700 \\
\centering Cell$_{15,22}$ & 500 \\ \bottomrule
\end{tabular}
\label{table3}
\end{table}

\begin{table}[!htb]\centering
\caption{Evacuation number at each exit.}
\resizebox{.3\columnwidth}{!}{
\begin{tabular}{lll}
\toprule
\textbf{Exits} & \textbf{Total evacuees} \\ \midrule
Exit 1 & 1600 \\
Exit 2 & 1700 \\
Exit 3 & 3200 \\
Exit 4 & 1700 \\
Total & 8200 \\ \bottomrule
\end{tabular}
}
\label{table4}
\end{table}

\begin{figure}[ht!]
\centering
\includegraphics[scale=0.3]{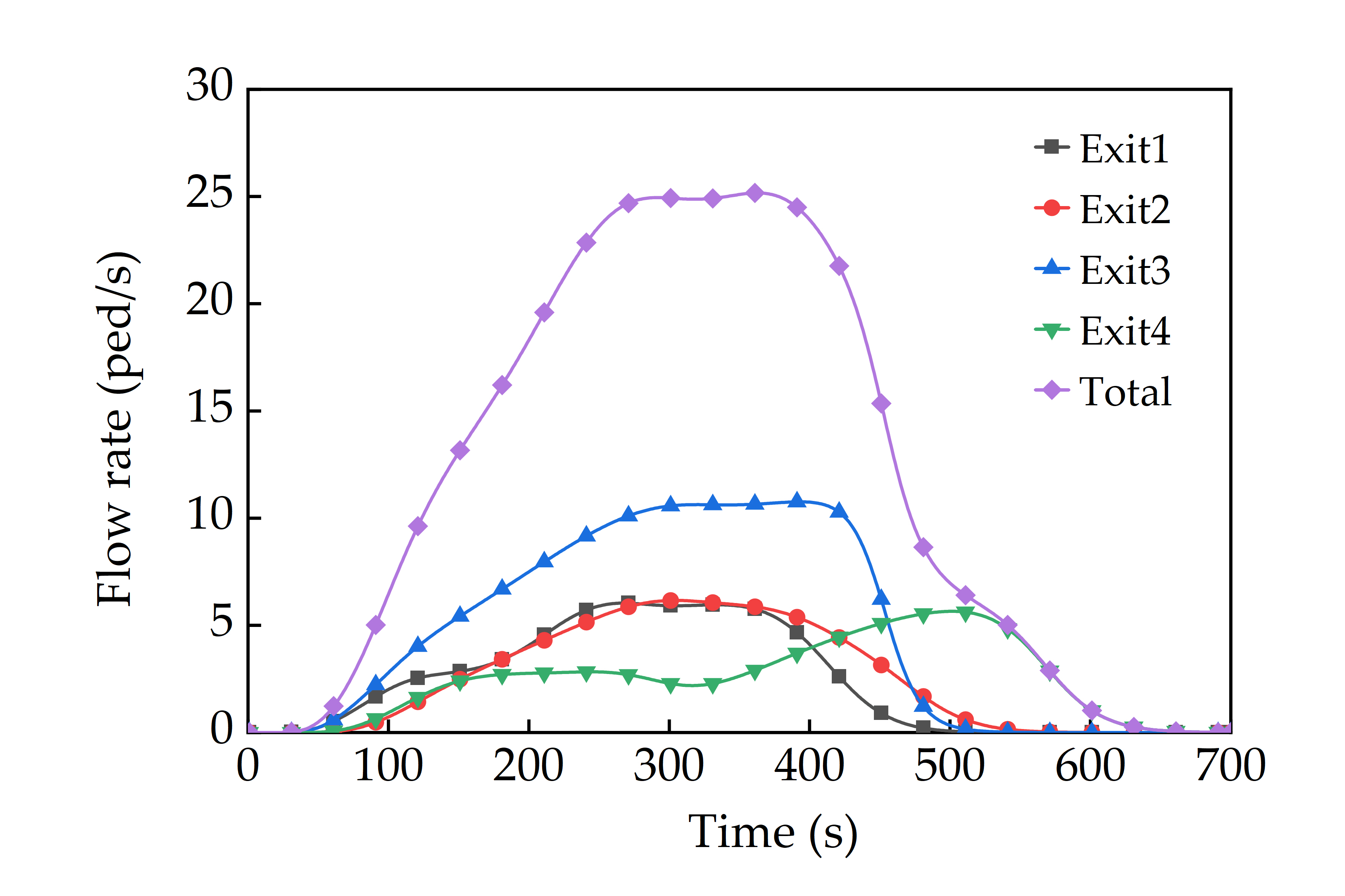}
\caption{Temporal evolution of flow rates at individual exits and the overall evacuation flow rate. 
\label{fig9}}
\end{figure}

\begin{figure}[ht!]
\centering
\includegraphics[scale=0.3]{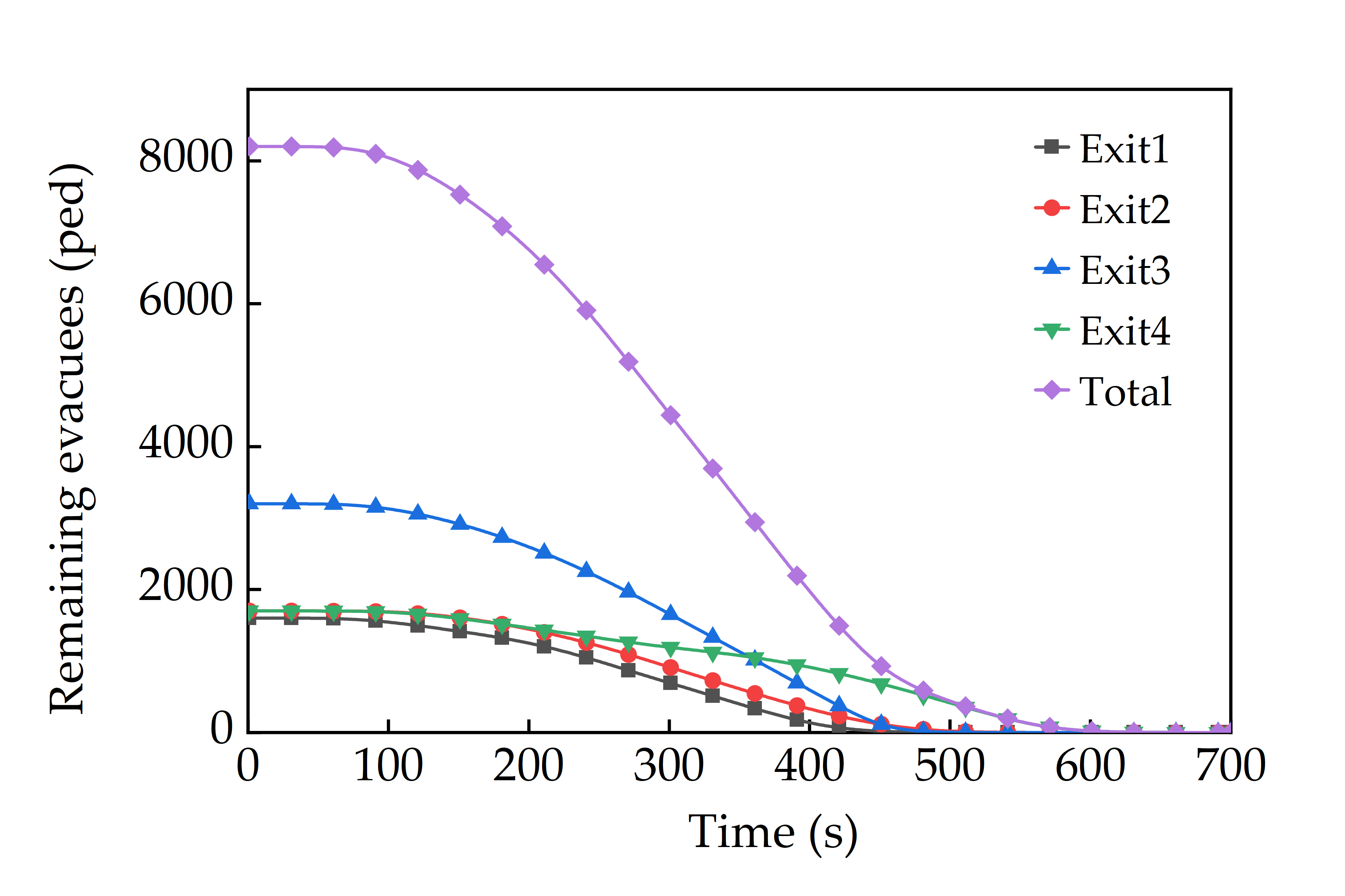}
\caption{Temporal evolution of both individual exits and the overall remaining evacuees. 
\label{fig10}}
\end{figure}

\begin{figure}[ht!]
\centering
\includegraphics[scale=0.25]{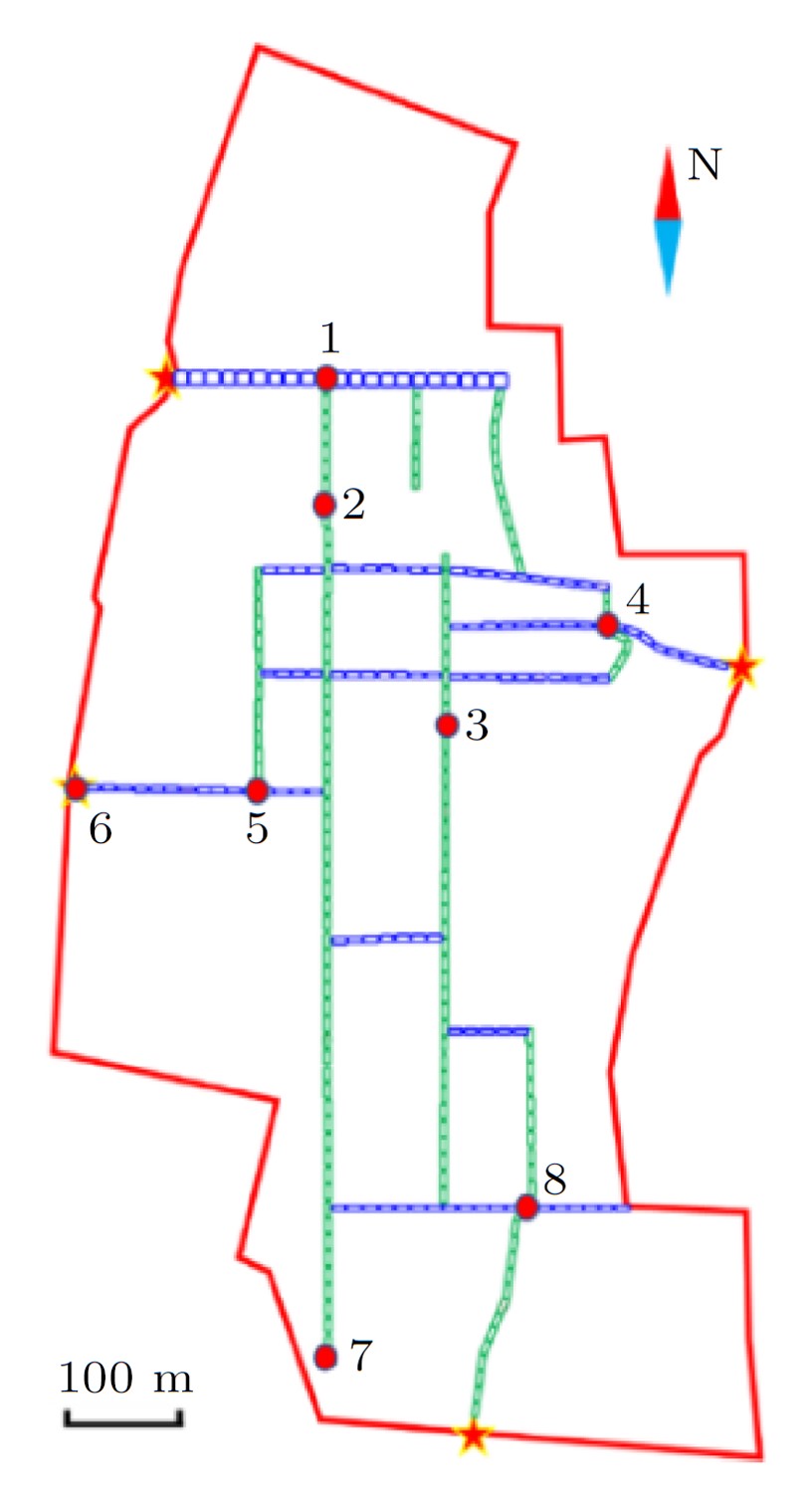}
\caption{Location distribution of observed cellular. 
\label{fig11}}
\end{figure}

\subsection{Results and disscussion\label{subsection4.2}}
Based on the provided cellular loading data and state update rules, extensive evacuation simulations of the campus were conducted using the mesoscopic cellular automata model.
According to the simulation results, the relationship between the evacuation counts at each exit and the total evacuation count over time is illustrated in Fig.\ref{fig9}. From the graph, it is evident that after 40 seconds, the curves exhibit an ascending trend, indicating that pedestrians begin to move from source-loaded cells to evacuate from the campus after the initial 40 seconds. The peak evacuation flow rate occurs between 250-400 seconds. Analyzing the curves depicting the relationship between the evacuation flow rate at each exit over time reveals that the peak evacuation flow rate for exits 3 and 4 occurs later than the corresponding peak evacuation flow rate for exits 1 and 2. This delay is attributed to the higher evacuation pressure on exit 3 and the greater distance of exit 4 from the loaded cells within the evacuation sub-network. Overall, the evacuation flow rate for the campus shows an initial increase, followed by stabilization, and then a decline.

Fig.\ref{fig10} illustrates the temporal evolution of the remaining evacuees for each exit and the overall. Generally, the curves exhibit a stable state during the early stages of evacuation, followed by a significant decline after a period of stability. In the later stages of evacuation, the decreasing trend begins to decelerate, and the remaining evacuees gradually approach zero. At the 672-second mark, the total remaining evacuees stand at 0.4925 individuals, indicating the complete evacuation of all personnel from the campus.

To gain further insights into the micro-level variations of individual cellular states, this study focuses on the observation of eight observed cells. The observation involves monitoring the temporal evolution of both the quantity and speeds within these cells. The specific locations of the designated observed cells are illustrated in Fig.\ref{fig11}. The chosen eight observed cells are strategically dispersed across four evacuation sub-networks, encompassing one source-loaded cell, one exit cell, four road junction cells, and two road cells.

\begin{figure}[ht!]
\centering
\includegraphics[scale=0.3]{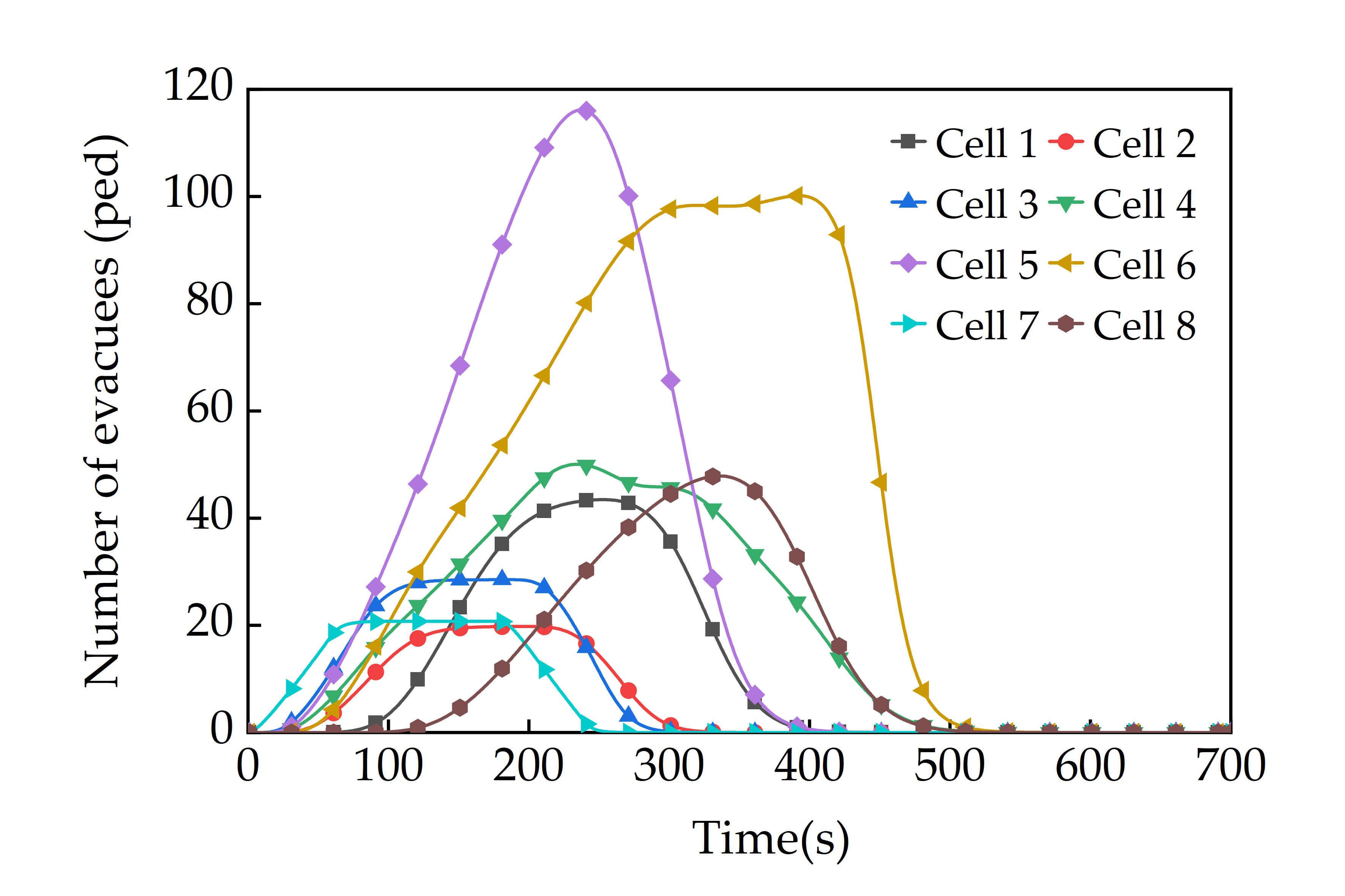}
\caption{Temporal evolution of evacuee quantities within the observed cell. 
\label{fig12}}
\end{figure}

\begin{figure}[ht!]
\centering
\includegraphics[scale=0.3]{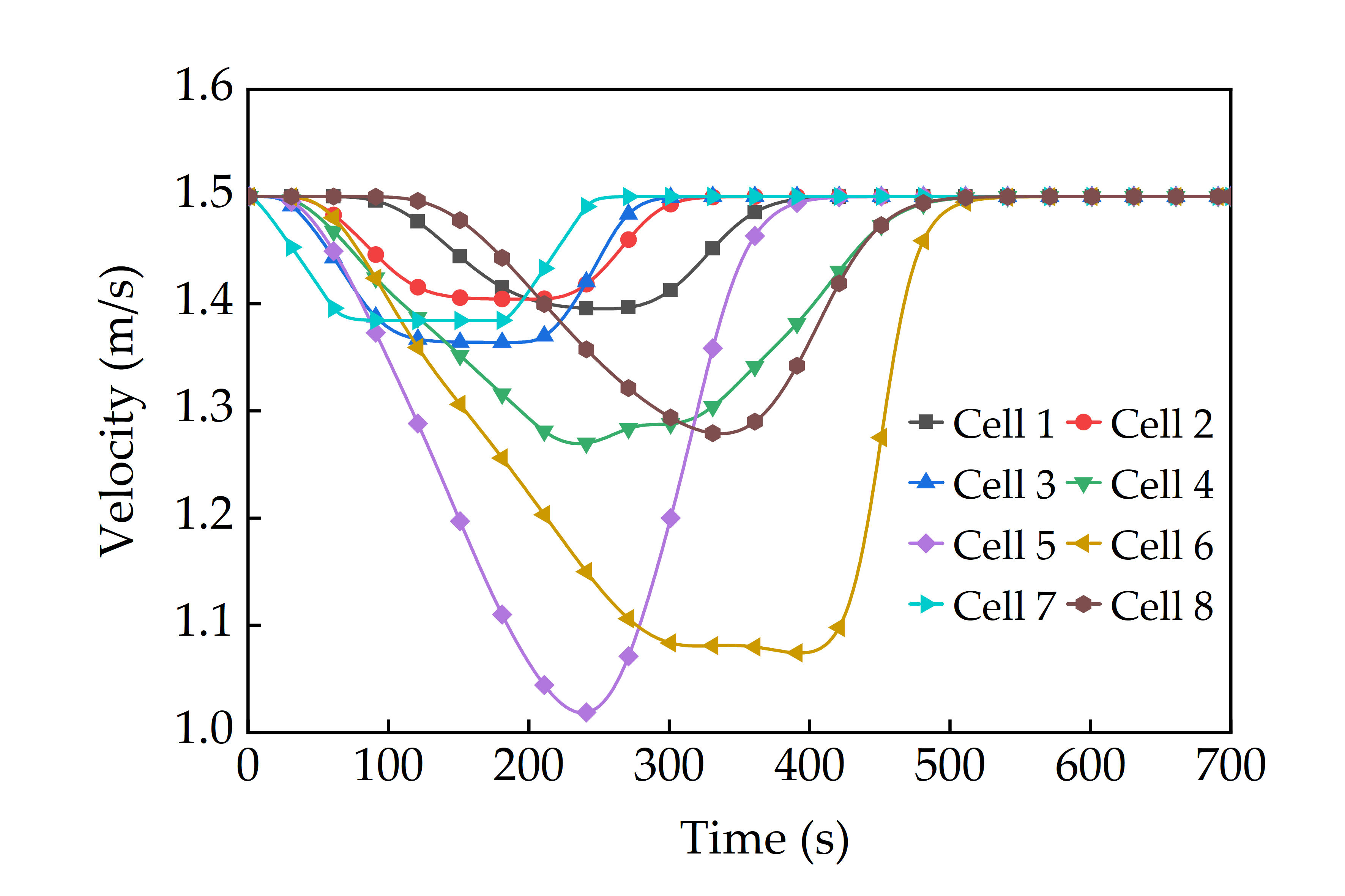}
\caption{Temporal evolution of evacuation speed within the observed cell. 
\label{fig13}}
\end{figure}

Figs.\ref{fig12} and \ref{fig13} depict the temporal variations in density and velocity within the selected eight cells. It is discernible from the graphs that Cell 6, designated as the exit cell, exhibits a peak density at 420 seconds, later than the peak times observed in other observed cells. Cell 5, positioned as the road node cell in front of Exit 3, bears a substantial evacuation pressure, with a peak density significantly higher than that of other observed cells, reaching nearly 120 individuals. Correspondingly, the velocity at its location is the lowest among the eight observed cells, slightly exceeding 1 m/s but remaining within the range of 1 m/s to 1.5 m/s, a fluctuation observed in the velocity of all observed cells during the evacuation process.

\begin{figure}[ht!]
\centering
\includegraphics[scale=0.55]{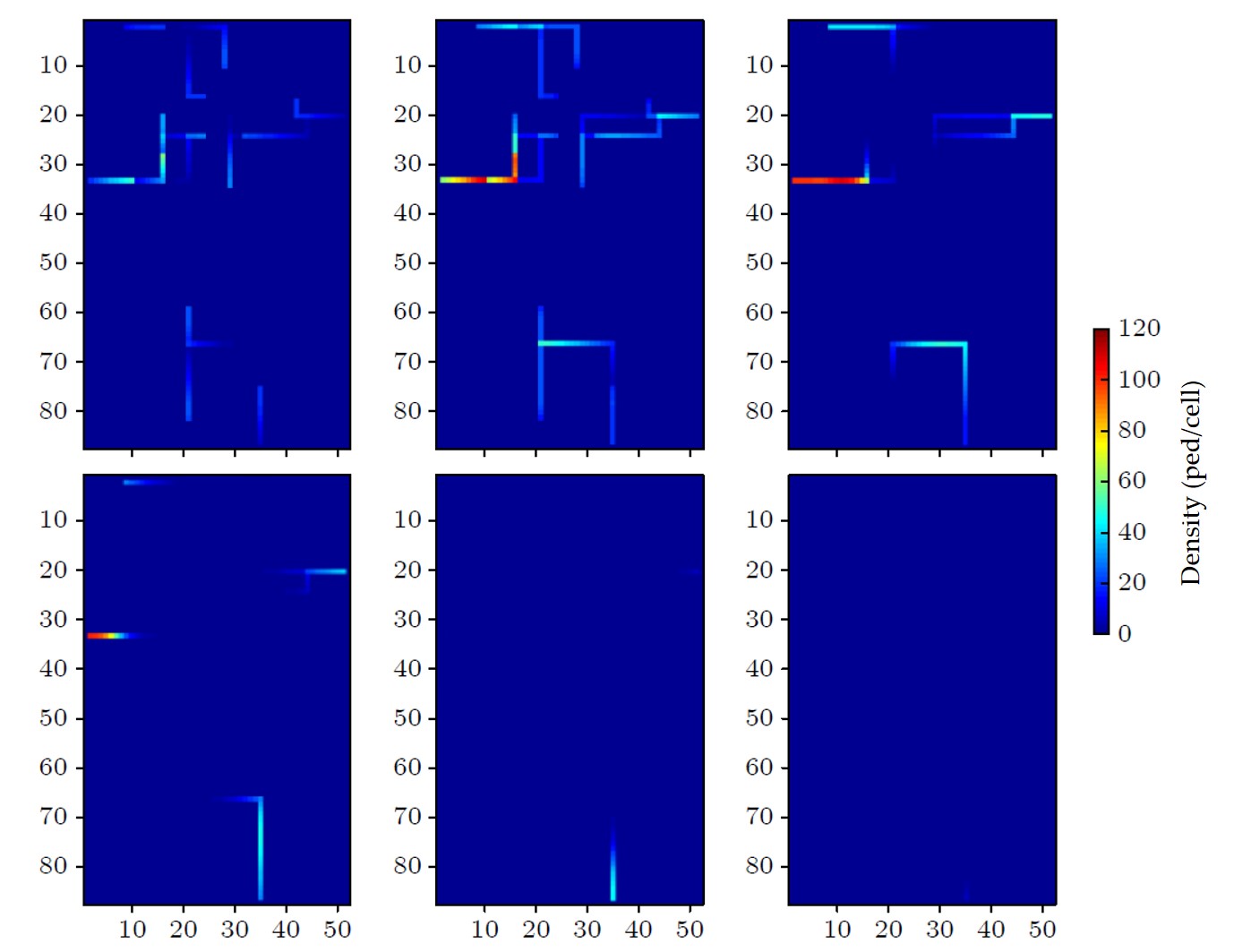}
\captionsetup{singlelinecheck=false, justification=centering}
\caption{Spatial staffing distribution:\\(a) \(t = 100 \, \text{s}\); (b) \(t = 200 \, \text{s}\); (c) \(t = 300 \, \text{s}\); (d) \(t = 400 \, \text{s}\); (e) \(t = 500 \, \text{s}\); (f) \(t = 600 \, \text{s}\).
\label{fig14}}
\end{figure}

Fig.\ref{fig14} illustrates the distribution of evacuees in each cell at time intervals of 100 seconds. Notably, between 200 s and 400 s, elevated evacuees are observed along various roads, signifying considerable evacuation pressure at each exit. At 500 seconds, Exit 1, Exit 2, and Exit 3 exhibit reduced populations in their exit cells, while Exit 4 still has pedestrians yet to complete evacuation. By 600 seconds, only a small number of individuals remain in the evacuation scene, signaling the nearing conclusion of the evacuation.

\section{Conclusions} \label{section5}

(1) This study introduces a novel mesoscopic cellular automata model for large-scale evacuation, utilizing road cells to establish a continuous model for simulating pedestrian evacuation. This addresses the challenge faced by traditional microscopic cellular automata models in simulating evacuation in large-scale spaces. The model is applied to simulate a university campus, yielding favorable simulation results.

(2) The total evacuation time in the simulation is 672 seconds, reaching the peak evacuation flow between 250-400 seconds. During this period, the peak occupancy at the  observed cell 5 approaches 120 evacuees. It is crucial to avoid excessively dense crowds during the evacuation process to prevent stampede incidents.

(3) Simulation reveals that within the evacuation sub-network around exit 3, there are numerous living and facility areas, resulting in the highest cell loading and the maximum evacuation pressure on the road network. Therefore, in practical evacuation considerations, diverting individuals in this area could alleviate localized road network evacuation pressure.

(4) Analysis of the simulation results indicates that exit 4 has a relatively small number of evacuees but requires the longest evacuation time among the four exits. This is attributed to the considerable distance between the loading area and the exit. Consequently, in practical evacuations, planning a shortcut between crowded areas and campus exits could expedite the evacuation process and reduce evacuation time.

\bibliographystyle{aasjournal}



\end{document}